\documentclass[12pt]{article}
\usepackage{amssymb}
\usepackage{epsf}
\usepackage{a4}

\newcommand{\btb}{\bar{\bt}}
\newcommand{\lmb}{\bar{\lm}}

\newcommand{\vb}{\bar{v}}

\newcommand{\vecx}{{\bf x}}       %(vette x)
\newcommand{\al}{\alpha}
\newcommand{\bt}{\beta}
\newcommand{\gm}{\gamma}
\newcommand{\dl}{\delta}

\newcommand{\et}{\eta}

\newcommand{\lm}{\lambda}
\newcommand{\rh}{\rho}

\newcommand{\ta}{\tau}

\newcommand{\ph}{\phi}
\newcommand{\vr}{\varphi}

\newcommand{\om}{\omega}

\newcommand{\Gm}{\Gamma}

\newcommand{\Sg}{\Sigma}

\newcommand{\half}{\frac{1}{2}}

\newcommand{\Tr}{\mbox{Tr}\,}

\newcommand{\dmu}{\partial_{\mu}}

\newcommand{\phd}{\ph^{\dagger}}

\newcommand{\eela}[1]{\label{#1}\end{equation}}
\newcommand{\eeala}[1]{\label{#1}\end{eqnarray}}
\newcommand{\be}{\begin{equation}}
\newcommand{\ee}{\end{equation}}
\newcommand{\bea}{\begin{eqnarray}}
\newcommand{\eea}{\end{eqnarray}}

\begin{document}
\title{
\vskip -100pt
{\begin{normalsize}
\mbox{} \hfill ITFA-97-02\\
\mbox{} \hfill  February 1997\\
\vskip  70pt
\end{normalsize}}
Numerical study of plasmon properties in the SU(2)-Higgs
model
% for $m_H \approx m_W$
}
\author{
Wai Hung Tang\thanks{email: tang@phys.uva.nl}\mbox{ }
and
Jan Smit\thanks{email: jsmit@phys.uva.nl}\mbox{ }
\\
Institute of Theoretical Physics, University of Amsterdam\\
Valckenierstraat 65, 1018 XE Amsterdam, the Netherlands
}
\date{revised version, July 1997}
\maketitle
\begin{abstract}
Using the (effective) classical approximation, we compute numerically
time-dependent correlation functions in the SU(2)-Higgs model around
the electroweak phase transition, for $m_H \approx m_W$.
The parameters of the classical model have been determined previously
by the dimensional reduction relations for time-independent correlators.
The $H$ and $W$ correlation functions correspond to gauge invariant fields.
They show damped oscillatory behavior from which we extract 
frequencies $\om$ and damping rates $\gm$. In the Higgs phase
the damping rates
have roughly the values obtained in analytic calculations in the 
quantum theory.
In the plasma phase (where analytic estimates for gauge invariant
fields are not available),
the damping rate associated with $H$ is an order of magnitude larger 
than in the Higgs phase, while the $W$ correlator appears to be 
overdamped, with a small rate. 
The frequency $\om_H$ shows a clear dip at the transition.
The results are approximately independent of the lattice spacing,
but this appears to be compatible with the lattice spacing dependence
expected from perturbation theory.
\end{abstract}

\clearpage

\section{Introduction}

The classical approximation has been introduced some time ago in
quantum field theory to avoid the difficulties of
numerical simulations of time dependent phenomena at high
temperature \cite{GrRu}. It is potentially a very powerful method for
nonequilibrium phenomena as well.
It has been used for the numerical study of the
Chern-Simons diffusion rate in the (1+1)D abelian Higgs model
%\cite{GrRuSha89,BochFo93,KraPo94,FoKraPo94,SmTa94,SmTa95},
[2--7], the (3+1)D SU(2)-Higgs model \cite{Amea,Krapriv,TaSm96} and pure
SU(2) gauge theory \cite{AmKra95}, as well as in a study of real
time properties of the electroweak phase transition
\cite{MoTu96}. The results have been encouraging. For example,
the sphaleron rate in the classical abelian Higgs model obtained by
numerical simulations agrees with the analytic expression of
quantum field theory
%\cite{KraPo94,FoKraPo94,SmTa94,SmTa95},
[4--7].
In the corresponding pure SU(2) and SU(2)-Higgs cases the rate
appeared to be lattice spacing independent in refs.\ \cite{AmKra95,TaSm96},
%
%On the other hand, the classical results for the rate are
%puzzling in the high temperature region of the abelian model
%\cite{SmTa95} and the analytical sphaleron rate in the Higgs phase of the
%SU(2)-Higgs model is in disagreement with results of
%recent numerical simulations
%\cite{Krapriv,TaSm96}. This may be due to bad ultraviolet
%properties of the observable used for estimating the Chern-Simons
%diffusion \cite{Krapriv}, suggesting a tricky situation where
%approximate lattice spacing independence is no guarantee for
%physical results. In the corresponding case of the topological
%susceptibility in euclidean lattice QCD such an unfavorable situation
%is indeed possible, see e.g.\ Fig.\ 7 in ref.\ \cite{LaSmVi90}.
%
but these results are now considered to be misleading, because
of bad ultraviolet properties of the observable
used for estimating the Chern-Simons diffusion. 
Recent simulations with improved observables have led to much better results 
in the Higgs phase and somewhat different values for the rate
in the plasma phase \cite{MoTu97,AmKra97}. The case of SU(3)
gauge theory has also been studied \cite{Mo97}. 
However, these new results for the rate show lattice spacing dependence.

A recent analysis of the dynamics of hot quantum
gauge fields led to the conclusion that the classical rate should 
depend on the lattice spacing \cite{ArSoYa97}. Other
lattice artefacts such as anisotropies have been anticipated earlier
\cite{BoMcLSm95}. These have been studied in detail in \cite{Ar97},
where a useful fudge factor was derived for converting the rate in the 
classical theory to an estimate for the rate in the quantum theory. 
These studies were analytic, based on a separation
of scales $g T$, $g^2 T$, in resummed perturbation
theory. This may be questionable when the gauge coupling $g$
is not very small. 
 
In any case, the very validity of the effectively\footnote{By
`classical' we shall always mean `effectively classical' in this paper,
i.e.\ in terms of a classical action with effective coupling constants.}
classical approximation for real time processes needs further
clarification.
For this reason we undertook the present
study of correlation functions in the classical SU(2)-Higgs theory.
The emphasis here is on time dependent correlation functions. The
static correlation functions, which describe
the initial conditions of the time dependent ones, have been well studied
since they correspond to a dimensional reduction approximation of the
quantum theory at high temperature. The static theory can be
renormalized, with finite renormalizations determined by matching the
classical to the quantum theory \cite{Kajea96a}.
An important question is if the time dependent theory is also renormalizable.

Recently
\cite{AaSm96}, time dependent correlation functions have been
studied in hot classical $\ph^4$ theory using perturbation theory
and it was shown that the ultraviolet divergencies can be
absorbed in counterterms, the same mass counterterms as needed in
the static theory. This led in particular to a finite classical
plasmon damping rate. Matching the parameters of the classical
static
theory with those in the quantum theory, this classical damping
rate was found to agree with the hard thermal loop expression of
the quantum theory.
Such matching was applied previously to the SU(2)-Higgs model for
the computation of the Chern-Simons diffusion rate \cite{TaSm96}.
The analysis is to be completed by a more detailed comparison between
the fully time dependent classical and quantum theories \cite{AaSm97}. 

Our approach is different from propositions based on
an effective theory of hard thermal loops \cite{BoMcLSm95,HuMu96},
in which the cutoff of the classical theory should not be removed. 
We are instead aiming at the construction of a classical SU(2)-Higgs
theory that is renormalizable (as in the scalar case \cite{AaSm96}),
such that physical results are free of regularization artefacts, 
for sufficiently large cutoff or small lattice spacing.
We do not know yet if such a construction exists. It may involve
nonlocal counterterms or other degrees of freedom, e.g.\ as in
\cite{Na94}.
The reason is that in gauge theories the divergences in the classical 
theory are non local,
of a form that is analogous to the hard thermal loops
in the quantum theory. Furthermore, the form of the usual classical
hamiltonian does not allow for a counterterm to compensate for the 
divergence in the Debije mass \cite{TaSm96}. 
Also the plasmon frequency is divergent \cite{Ar97}. 
On the other hand, in the static case,
the large Debije mass is the rationale for 
integrating out the corresponding modes to get a simpler form of
the dimensional reduction approximation. Similarly, in the time
dependent case, without additional nonlocal counterterms, 
any modes with divergent frequencies or damping
rates may simply decouple from coarse grained large time behavior.
The question is, if what is left over can be interpreted as a
(nonperturbatively) renormalizable theory.

In this paper we deal with the usual classical model and
concentrate on relatively
simple time dependent correlation functions. First of all,
we want to see if it is feasible to compute such autocorrelation
functions numerically with reasonable accuracy, find out how the
results depend on the lattice spacing, using
the counterterms known from the static theory, and
compare with hard thermal loop results of the quantum theory.

Such a comparison is not straightforward in the high temperature phase.
In the analytic calculations the correlation functions are constructed
from the elementary field variables using gauge fixing, while
nonperturbatively one tends to use gauge invariant composite local
fields. Even if the final results obtained analytically are
gauge invariant, it is not clear that these refer to
the same quantities as those obtained with the gauge invariant fields.
For example, in the plasma phase
there is a large discrepancy between the two cases
in the screening masses with $W$ quantum numbers
%\cite{BuPhi95,Guea96,Foea95,Kajea96b,Phiea96,Karea96}
[24-29] 
(see also the reviews \cite{Ja95,Rum96} and the interpretation
given in ref.\ \cite{BuPhi97}).
In the Higgs phase the results agree, as can be understood from
the usual argument which we now review.

The simplest gauge invariant composite fields (which we also
use) read in the continuum
\bea
H &=& \vr^{\dagger}\vr,\;\;\; \vr = \left(\begin{array}{c} \vr_u\\ \vr_d
\end{array} \right), \\
W_{\mu}^{\al}&=& i \Tr \phd
(\dmu - iA_{\mu}^{\bt}\, t_{\bt})
\ph\, t_{\al},\;\;\;
\ph = \left(\begin{array}{cc}\vr_d^* & \vr_u\\ -\vr_u^* & \vr_d
\end{array}\right),
\eea
with $\vr$ the Higgs doublet,
$A_{\mu}^{\al}$ the gauge fields  and $t_{\al}
= \tau_{\al}/2$ the SU(2) generators. In the Higgs phase
these fields correspond to the usual Higgs and gauge fields.
Since $\ph=\sqrt{H}\, V$ with $V\in SU(2)$, $W_{\mu}^{\al}$ is
essentially the gauge field in the unitary gauge $V=1$, as
can be seen in perturbation theory by replacing $\vr$ by its vacuum
expectation value, neglecting fluctuations.
This is not useful in the plasma phase where the theory
behaves like QCD and perturbation theory runs into infrared
problems. Nonperturbatively $H$ and $W_{m}^{\al}$ ($m=1,2,3$) are
primarily characterized by their behavior under spatial rotations
and weak isospin: (scalar,scalar) and (vector,vector),
respectively. We have not studied the time component $W^{\al}_0$,
which transforms as (scalar,vector).

To get an idea of the properties of the system being simulated we
also computed static correlators of $H$ and $W$ and estimated
their screening masses. The temperatures varied around the
electroweak transition and parameters were furthermore chosen
such that for the zero temperature masses $m_H \approx m_W$. From the
time dependent correlation functions we then attempted to extract
plasmon frequencies and damping rates of $H$ and $W$.

We should warn the reader about our terminology.
For convenience we often call in this paper the frequencies and
attenuation rates associated with the composite $W$ and $H$ fields
at zero momentum: `plasmon' frequencies and damping rates. 
This is controversial, since the terminology `plasmon' is 
normally associated with the collective modes of the gauge field
$A_{\mu}$. 
%(It is not even clear to us that what we call
%$H,W$ plasmon properties are determined solely by the $H,W$ quantum numbers
%(as is the case for zero temperature where particle masses
%and widths are independent of the `interpolating fields'), 
%and not also by the specific $H,W$ fields realizing the
%$H,W$ quantum numbers. In the time-independent case this ambiguity
%does not arise, because the screening masses can be related to the 
%eigenvalues of a transfer operator acting in a time slice.) 

\section{Classical SU(2)-Higgs model at finite temperature}

We use the same notation as in \cite{TaSm96}.  The classical
SU(2)-Higgs model is defined on a spatial lattice with lattice
distance $a$. The parallel transporters (lattice gauge field) are
denoted by $U_{m\vecx}$, $D_m\bar\vr_{\vecx}=
U_{m\vecx}\bar\vr_{\vecx+\hat{m}} - \bar\vr_{\vecx}$ is the covariant
lattice derivative acting on the lattice Higgs doublet
$\bar\vr_{\vecx}$, $U_{mn\vecx}$ is the product of parallel
transporters around the plaquette $(\vecx mn)$. The canonical momenta
in the temporal gauge are denoted by $\bar\pi$, $\bar\pi^{\dagger}$
and $\bar E^{\al}_m$, with nontrivial Poisson brackets
\be
\{U_{m\vecx},\bar{E}^{\al}_{m\vecx}\} = iU_{m\vecx} \ta_{\al},
\;\;\;
\{\bar{\vr}_{\vecx},\bar{\pi}^{\dagger}_{\vecx}\} = 1.
\ee
The effective classical hamiltonian is given by
\bea
\frac{H_{\rm eff}}{T} &=& \btb \bar{H},\\
\bar{H} &=& \sum_{\vecx}
[\half\,z_E \bar{E}^{\al}_{m\vecx} \bar{E}^{\al}_{m\vecx}
+ z_{\pi}\bar{\pi}^{\dagger}_{\vecx}\bar{\pi}_{\vecx}
+ \sum_{m<n} (1-\half\, \Tr U_{mn\vecx})
\nonumber\\ &&\mbox{}
+ (D_m\bar{\vr}_{\vecx})^{\dagger}D_m\bar{\vr}_{\vecx}
+ \lmb(\bar{\vr}^{\dagger}_{\vecx}\bar{\vr}_{\vecx} - \vb^2)^2],
\eea
where $T$ is the temperature and $\bar\bt$, $\bar\lm$ and $\vb^2$ are
effective couplings which depend on $a$, $T$ and the couplings in the
4D quantum theory,
\bea
\bar\bt &\approx& \frac{4}{g^2 aT},\;\;\;
\bar \lm \approx \frac{m_H^2}{2m_W^2},\\
4\lmb\vb^2 &=& a^2 T^2 \left[ \frac{m_H^2}{T^2}
-
\left(\frac{3}{2}\,g\right)^2\,
\left(
\frac{3+\rh}{18} - \frac{3+\rh}{3}\, \frac{2\Sg}{aT}\,
\right)\,
\right.
\label{VBT}\\
&& \mbox{} \left.
-
\left(\frac{3}{2}\,g\right)^4
\frac{1}{8\pi^2}\, \left(
\frac{149+9\rh}{486}
 + \frac{27 + 6\rh -\rh^2}{27}\,
\ln\frac{aT}{2} - \frac{27\et + 6\rh\bar{\et} - \rh^2
\tilde{\et}}{27} \right)\right],
\nonumber\\
\Sg &=& 0.252731.
\eea
Here $\rh = m_H^2/m_W^2$ and $g$ and $m_H$ are parameters in the
MS-bar scheme at scale $\mu_T \equiv 4\pi\exp(-\gm_E) T\approx
7T$. These relations follow from matching the
static
classical theory to the
dimensionally reduced quantum theory \cite{TaSm96}, using the results
of ref.~\cite{Faea2}.
As in \cite{TaSm96} we set $z_{E} =  z_{\pi} = 1$.
The implementation of these
parameters has to wait for a detailed matching between the
time dependent classical and quantum theories.

The equations of motion follow from the above hamiltonian and Poisson
brackets. The initial conditions are distributed according to the
classical partition function
\be
Z = \int DE DU D\pi D\vr\, [\prod_{\vecx\al}
\dl(G_{\vecx}^{\al})] \, \exp(-H_{\rm eff}/T),
\label{8}
\ee
where $\dl(G_{\vecx}^{\al})$ enforces the Gauss constraint, which is
part of the temporal gauge formalism. Time dependent correlation
functions of observables $O$ are defined by averaging $O(t)O(0)$ over
initial configurations,
\be
\langle O(t) O(0) \rangle = \frac{1}{Z} \int DE DU D\pi
D\vr\, [\prod_{\vecx\al} \dl(G_{\vecx}^{\al})] \,
\exp(-H_{\rm eff}/T)\, O(t) O(0).
\ee
Examples of such observables are the simple gauge invariant fields we
use to study the $W$ and Higgs excitations:
\bea
W^{\al}_{m\vecx} &=& i\Tr \ph^{\dagger}_{\vecx} U_{m\vecx}
\ph_{\vecx + \hat m} \ta_{\al},
%\;\;\; \ph = (i\ta_2\bar\vr^*,\bar\vr),
\label{OW}\\
H_{\vecx} &=& \bar\vr^{\dagger}_{\vecx} \bar\vr_{\vecx}.
\label{OH}
\eea

The static system (i.e.~the system described by the classical
partition function at time zero) is believed to be well understood
because of its connection with dimensional reduction. For given
$\vb^2$ and relatively small $\lmb$ there is a critical value
of $\bar\bt$, above which the system is in the Higgs phase and below
which it is in the plasma phase. The transition is of first order
\cite{Faea2}, weakening in strength as $m_H/m_W$ increases, and
turning into a crossover near $m_H \approx m_W$ \cite{crossover}.

As in \cite{TaSm96}, we have chosen $\bar\lm=1/2$, which means
$m_H/m_W \approx 1$. The transition is then still clearly visible,
especially in the rate of Chern-Simons diffusion
\cite{TaSm96}. Accordingly set $\rh = 1$ in the application of
eq.~(\ref{VBT}). We also neglect the running of $g$ and $m_H$ with
temperature, and choose $g = 2/3$, which implies $aT = 9/\bar\bt$. For
given $\vb^2$ eq.~(\ref{VBT}) then gives $T/m_H$ as a function of
$\bar\bt$.

Keeping instead $T/m_H$ fixed while increasing $\btb$ and changing
$\vb^2$ according to (\ref{VBT}) should get us closer to the continuum
limit. Physical quantities of the static theory then become $\btb$
independent. It is not clear at this point if the same holds for
physical quantities related to the full time dependent theory, such as
plasmon masses and rates.

\begin{table}[t]
\centerline{\begin{tabular}{||c|r|l|l|l||} \hline
\multicolumn{1}{||c|}{$\bar{\beta}_c$} &
\multicolumn{1}{c|}{N} &
\multicolumn{1}{c|}{$\beta^c_H$} &
\multicolumn{1}{c|}{$\bar{v}^2_c$} &
\multicolumn{1}{c||}{$T_c/m_H$}
\\ \hline
12 & 20 & 0.347733 & 0.26295 & 2.14 \\ \hline
12 & 24 & 0.34772 & 0.26273 & 2.15 \\ \hline
20 & 32 & 0.34173 & 0.15597 & 2.15 \\ \hline
\end{tabular}}
\caption{
Critical values of $(\btb,\vb)$ for $\lmb=1/2$. The values of $T_c/m_H$
were calculated using eq.\ \protect (\ref{VBT}).
}
\label{v2table}
\end{table}
Table \ref{v2table} shows some pseudo critical values $\btb_c$ and
$\vb^2_c$ on lattices of size $N^3$.  These values are obtained by a
translation of the dimensional reduction results of \cite{Faea2} into
the parameterization used here in terms of $\lmb$ and $\vb^2$.  For
the record we mention the connection with the parameterization used in
\cite{Faea2}: $\btb = \bt_G$, $\lmb = 4\bt_G\bt_R/\bt_H^2$,
$(\vb^2/\lmb) = 4(3\bt_H - 2\bt_R -1)/\bt_H$, $\bt^A_2 = \bt^A_4 = 0$.
The last two parameters were not zero in the action used in
\cite{Faea2}, but the difference does not seem to be important for the
location of the phase transition \cite{TaSm96}.

\section{Numerical computation}

We use the algorithm and numerical implementation offered in
ref.~\cite{Kra95}.
A simulation consisted of alternating Langevin runs (the generation of
initial conditions) and Hamilton runs (integration of the equations of
motion). Spatial correlation functions used for the study of screening
properties were averaged over the Langevin runs. The Langevin runs
lasted typically 120 (in lattice units), which we found sufficient to
decorrelate observables such as the `link' $\Tr \phd U \ph$,
while the Hamilton runs lasted in most cases several thousand.
Details on the statistics will be given below where we
summarize the parameter values in Table \ref{partable}.

The spatial correlation in the 3-direction at distance $d$, of a local
observable $O_{\vecx}$ at zero transverse momentum, was estimated as
follows,
\bea
G_O(d) &=& \frac{1}{N}\sum_{x_3}\left[
\overline{\tilde{O}_{x_3 + d} \tilde{O}_{x_3}}
-\overline{\tilde{O}_{x_3 + d}}\;\;\overline{\tilde{O}_{x_3}}
\right],
\label{Gdef}\\
\tilde{O}_{x_3} &=& \frac{1}{\sqrt{N^2}}\sum_{x_1,x_2} O_{x_1,x_2,x_3},
\eea
where the bar denotes the average over the Langevin bins (we used
periodic spatial boundary conditions). In the limit of an infinite
number of bins $\overline{O_{\vecx}} \to \langle O_{\vecx} \rangle$
which is independent of $\vecx$. For the Higgs mode we used
$O_{\vecx}= H_{\vecx}$ (cf.~(\ref{OH}) and we averaged the
correlators over the three different directions (3, 1, and 2). For the
$W$-mode we used the transverse components $O_{\vecx} = \sum_{\al=1}^3
W^{\al}_{k\vecx}$, $k=1,2$, which leads to a $G_{W_k}(d)$ independent
of $k$, and we averaged over $k$, $G_W(d) = [G_{W_1}(d) +
G_{W_2}(d)]/2$, and over the three different directions. (The
summation over $\al=1,2,3$ picks out a particular direction in isospin
space; we could have improved statistics a little by correlating and
averaging the $\al$'s as well).

For the time dependent correlators we used microcanonical averaging
over the Hamilton parts of the simulation in addition to canonical
averaging over initial conditions at the end of the Langevin
parts. The autocorrelation functions were constructed analogously to
(\ref{Gdef}),
\be
C_O(t) =
\frac{1}{t_b}\, \int_0^{t_b} dt_0\,
\left[ \overline{O(t_0+t) O(t_0)}
- \overline{O(t_0+t)}\;\;\overline{O(t_0)}\right],
\label{micro}
\ee
where $0 < t < t_{\rm run} - t_b$. Here the bar denotes the canonical
average and $t_{\rm run}$ is the maximum time in the Hamilton
run. When the number of initial conditions goes to infinity this
expression approaches the exact $\langle O(t) O(0) \rangle - \langle O
\rangle^2$. For the determination of the plasmon properties we used
for $O$ the zero momentum projections
\be
%W_k^{\al} = \frac{1}{\sqrt{N^3}}\sum_{\vecx} W_{k\vecx}^{\al},\;\;\;
W_k = \frac{1}{\sqrt{N^3}}\sum_{\vecx\al} W_{k\vecx}^{\al},\;\;\;
H = \frac{1}{\sqrt{N^3}}\sum_{\vecx} H_{\vecx},
\ee
and for $W$ we averaged over $k$.
% and summed over $\al$.

\section{Results for the screening correlators}

\begin{table}[t]
\centerline{
\begin{tabular}{||c|c|c|c|c|r|r|r||} \hline
\multicolumn{1}{||c|}{$\bar{\beta}$} &
\multicolumn{1}{c|}{$N$} &
\multicolumn{1}{c|}{$\vb^2$} &
\multicolumn{1}{c|}{$T/m_H$} &
\multicolumn{1}{c|}{$T/T_c$} &
\multicolumn{1}{c|}{$t_{\rm run}$} &
\multicolumn{1}{c|}{nc(sc)} &
\multicolumn{1}{c||}{nc(pl)} \\ \hline
%4 & 24 & 0.263 & - & - & $400$ & 25  & 17 \\ \hline
20 & 24 & 0.263 	& 0.88 	& 0.41 	& $1000$ & 15  & 15 \\ \hline
14 & 24 & 0.263 	& 1.59 	& 0.74 	& $6400$ &  15 & 18 \\ \hline
21.7 & 32 & 0.156 	& 1.67 	& 0.78 	&$7000$ & 31 & 12 \\ \hline
13 & 24 & 0.263 	& 1.82 	& 0.85 	& $5000$ &  42 & 42 \\ \hline
$13'$   & 20 & 0.263 	& 1.81 	& 0.85 	&$5000$ & 40 & 11 \\ \hline
12.5 & 24 & 0.263 	& 1.98 	& 0.92 	& $6400$ &  35 & 30 \\ \hline
12 & 24 & 0.263 	& 2.15 	& 1 	& $1500$  &  30 & 29 \\ \hline
11 & 24 & 0.263 	& 2.64 	& 1.23  & $400$ & 89  & 86 \\ \hline
$11'$  & 20 & 0.263 	& 2.63 	& 1.23 	&$2500$ & 57 & 56 \\ \hline
$11^\ast$ & 20 & 0.246 	& 3.26 	& 1.52 	&$3000$ & 61 & 61 \\ \hline
$18.3$ & 32 & 0.156     & 3.26 	& 1.52 	&$2500$ & 37 & 21 \\  \hline
10 & 24 & 0.263 	& 3.46 	& 1.61	& $1280$ & 24  & 48 \\ \hline
8 & 24 & 0.263 		& - & -  & $320$ & 41  & 51 \\ \hline
6 & 24 & 0.263 		& - & -  & $640$ & 35  & 22 \\ \hline
\end{tabular}}
\caption{
Summary of parameter values and statistics, 
ordered according to increasing $T/T_c$. 
The prime or $\mbox{}^\ast$ on $\btb$ distinguishes different $N$ and $\vb^2$.
}
\label{partable}
\end{table}

Table \ref{partable} gives a summary of the parameter values used in
our simulations.
The values of $T/m_H$ correspond to eq.\ (\ref{VBT}). (For $\btb=6$, 8 the
lattice spacing is so large that eq.\ (\ref{VBT}) breaks
down; note that $O(a)$ terms are neglected in $[\cdots]$.)
The ratios $T/T_c$ have been added for convenience and follow simply by
devision by $T_c/M_H(\mu_{T_c})$ in Table \ref{v2table}, i.e.\ without taking
into account the running of $m_H(\mu_T)$ with $T$.
The time $t_{\rm run}$ is the length of the Hamilton runs used in
the computation of the autocorrelation functions, in lattice
units; nc(sc) and nc(pl) are the number configurations (number
of Langevin runs) used in the calculation of the screening and
plasmon properties, respectively.

We first made a scanning simulation on the $N=24$ lattice with varying $\btb$
and $\vb^2$ fixed at 0.263, its value for $\btb_c=12$. This means
the temperature varied as shown in Table \ref{partable}. By fitting
the correlation functions $G_{H,W}(d)$ to $A\cosh[am(d-N/2)]$ we obtained
the screening masses $m$ shown in Table \ref{tscreen}.
\begin{table}[t]
\centerline{
\begin{tabular}{||c|c|c|c|c|c|c||} \hline
\multicolumn{1}{||c|}{$\bar{\beta}$} &
\multicolumn{1}{c|}{$N$} &
\multicolumn{1}{c|}{$T/T_c$} &
\multicolumn{1}{c|}{$am_{H}$} &
\multicolumn{1}{c|}{$am_{W}$} &
\multicolumn{1}{c|}{$m_{H}/g^2 T$} &
\multicolumn{1}{c||}{$m_{W}/g^2 T$}
\\ \hline
%4 & 3(2) & 0.8(30)      & 3(2) & 0.8(30) \\ \hline
20 &24&0.41& 0.59(5) & 0.50(2)   	&   2.95(25)  	&  2.5(1)  \\ \hline
14 &24&0.74& 0.34(1) & 0.41(2)   	&   1.19(4)   	&  1.44(7) \\ \hline
21.7 &32&0.78& 0.22(2) & 0.25(2)  	&   1.19(11)  	& 1.36(11) \\ \hline
13 &24&0.85& 0.28(3) & 0.35(2)   	&   0.91(9)   	&  1.14(7) \\ \hline
$13'$ &20&0.85& 0.26(2) & 0.39(3)	&   0.85(7)   	& 1.27(10) \\ \hline
12.5 &24&0.92& 0.22(3) & 0.38(1) 	&   0.69(9)   	&  1.19(3) \\ \hline
12 &24& 1& 0.146(29) & 0.37(2) 	&   0.44(9)   	&  1.11(6) \\ \hline
11 &24&1.23& 0.46(3) & 0.88(14)  	&   1.27(8)   	&  2.4(4)  \\ \hline
$11^\prime$&20&1.23&0.44(3)&0.96(11)&  1.21(8) 	& 2.64(30) \\ \hline
$11^\ast$&20&1.52&0.59(5)&1.06(20) &   1.62(14) 	& 2.92(55) \\ \hline
18.3 &32&1.52& 0.31(3) & 0.65(10)	&   1.42(14) 	& 2.97(46) \\ \hline
10 &24&1.61& 0.68(4) & 1.34(28)  	&   1.7(1)    	&  3.4(7)  \\ \hline
8 &24& - & 1.12(20) & 1.04(58)     &   2.24(40)    &  2.1(12) \\ \hline
6 &24& - & 1.2(6) & 1.5(18)        &   1.8(9)      &  2.3(27) \\ \hline
\end{tabular}}
\caption{
Results for the screening masses.
}
\label{tscreen}
\end{table}

We made a least
squares fit to the data at $d=3$ -- 5, which led to reasonable
looking fits, and the errors were obtained with the jackknife method.
This procedure is of course not state of the art as in \cite{Phiea96},
but we only wanted to
get an overall impression of the data at the chosen parameters
using a moderate amount of computational effort.

After this scan we decided to
concentrate on two values of $\btb$ corresponding to two values of the
lattice spacing, in the Higgs phase as well as in the plasma phase. 
The parameters $\vb^2$ and $N$ were changed accordingly such
that the two $\btb$ values tentatively describe the same physical
situation. The guiding parameter values were those of the transition
obtained nonperturbatively in \cite{Faea2}, which are shown in Table
\ref{v2table}: $(\btb,N,\vb^2)$ = (12, 20, 0.263) and (20, 32,
0.156). Modulo scaling violations this corresponds to a ratio of
lattice spacings $12/20 = 0.6$ at approximately the same physical
volume ($20/32=0.625$). The new parameters in the Higgs phase were
obtained by changing $\btb=12 \to 13$ and $\btb=20 \to 21.7 \approx
13/0.6$, keeping fixed the corresponding $N$ and $\vb^2$. These pairs
of $(\btb,N,\vb^2)$ values correspond roughly to the same $T/M_H$
(using the matching formula (\ref{VBT}), as shown in Table
\ref{partable} (cf.\ $\btb = 13'$ and 21.7).
The same parameter combinations were used in
\cite{TaSm96} for the computation of the Chern-Simons diffusion
rate. 
Similarly the plasma phase parameters were obtained by
changing $\btb = 12 \to 11$ and $\btb=20 \to 18.3 \approx 11/0.6$,
again without changing the corresponding $N$ and $\vb^2$. However,
here the matching formula (\ref{VBT}) gives a larger mismatch in
$T/m_H$ (cf.\ $\btb = 11'$, 18.3 in
Table \ref{partable}). To keep $T/m_H$ fixed more
precisely we have to adjust $\vb^2$. Therefore we also did
measurements in the plasma phase at $(\btb,N,\vb^2)
= (11, 20, 0.246)$, for which $\vb^2$ was chosen such that $T/m_H$
equals its value at $(\btb,N,\vb^2) = (18.3, 32, 0.156)$.
In the Tables the $\btb$ value is denoted by $11^*$.

\begin{figure}
\epsfxsize 150mm
\centerline{\epsfbox{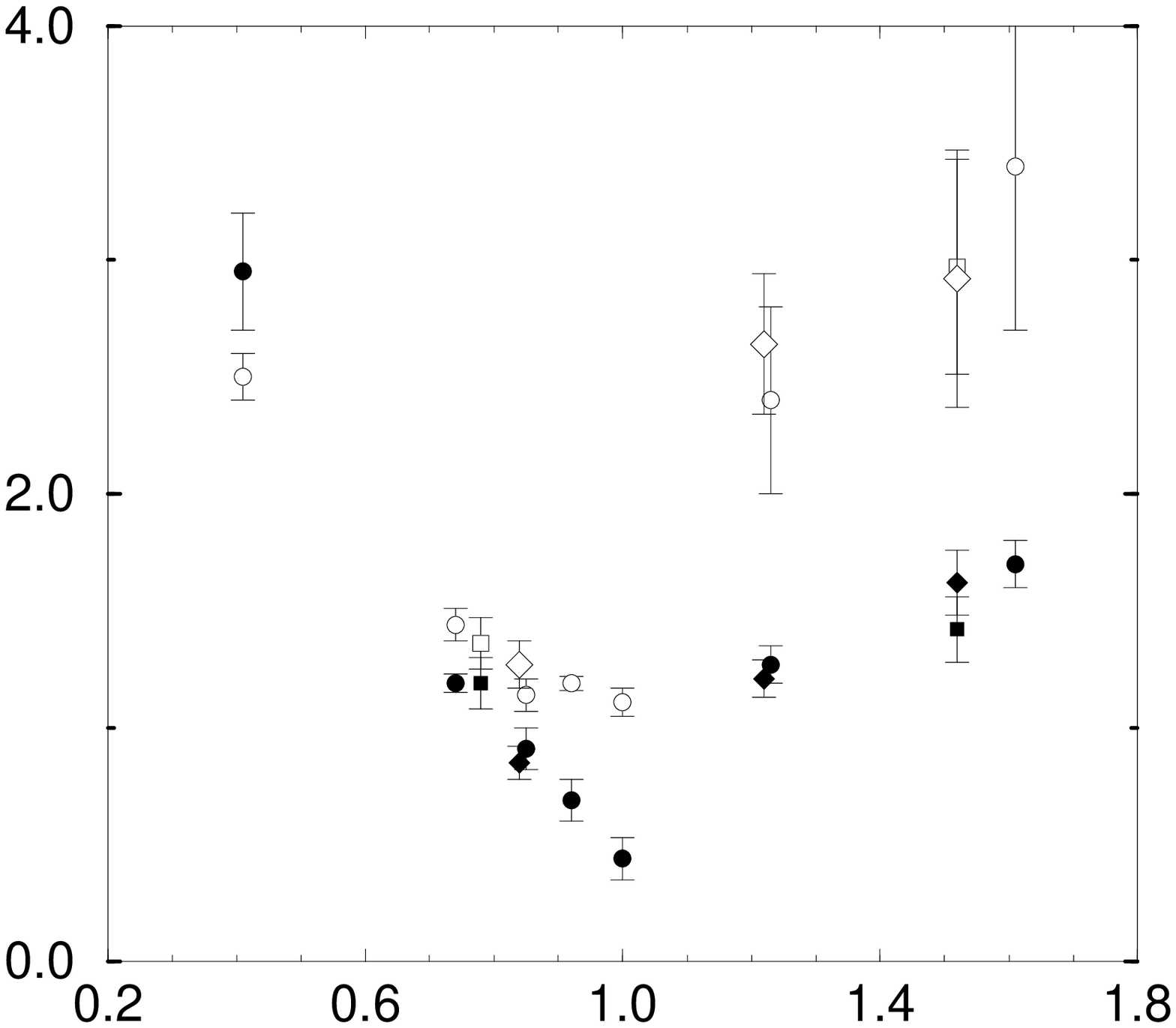}}
\caption{Screening masses $m_{H,W}/g^2 T$
versus $T/T_c$. Solid symbols correspond to $H$, open symbols to $W$;
circle: $24^3$ data; diamond: $20^3$ data; square: $32^3$ data.
}
\label{fscreen}
\end{figure}

Table \ref{tscreen} also shows the results for the screening masses of
these additional simulations.
We fitted the data at $d=3$ -- 5 for $N=20$ and at $d=4$ -- 7 for $N=32$.
Fig.\ \ref{fscreen} shows the screening
masses in physical units, $m/g^2 T$, versus $T/T_c$. The transition at
$T= T_c$ is clearly visible and we also see the ratio $m_H/m_W$
approaching 1 at low temperatures, as expected from the choice $\lmb =
1/2$.  (The dimensional reduction approximation is of course
questionable at the most left points where $T/M_H$ is only 0.88.)
We observe no significant volume dependence away from the phase
transition in the data at $T/T_c =0.85$ and 1.23.
The data at $T/T_c = 1.52$ 
($20^3$, $\btb=11^*$ and $32^3$, 18.3) show reasonable lattice
spacing independence (recall that the ratio of lattice spacings is 0.6).
The difference in $m/g^2 T$ between $\btb=13'$ ($20^3$) and 21.7 ($32^3$)
is evidently due to the difference in physical temperatures, $T/T_c = 0.84$
and 0.78, and not a signal of lattice spacing dependence.

\begin{table}[ht]
\centerline{
\begin{tabular}{||c|c|c|c|c||} \hline
\multicolumn{1}{||c|}{$\bar{\beta}$} &
\multicolumn{1}{c|}{$am_{{\rm eff}\,H}$} &
\multicolumn{1}{c|}{$am_{{\rm eff}\,W}$} &
\multicolumn{1}{c|}{$m_{{\rm eff}\,H}/g^2 T$} &
\multicolumn{1}{c|}{$m_{{\rm eff}\,W}/g^2 T$}
\\ \hline
$13^\prime$   & 0.24(2) & 0.43(4) & 0.78(7)   & 1.40(13) \\ \hline
21.7          & 0.18(3) & 0.27(3) & 0.98(16)  & 1.46(16) \\ \hline
$11^\prime$   & 0.64(5) & 1.65(14)& 1.76(14)  & 4.5(4)   \\ \hline
$11^\ast$     & 0.64(4) & 1.84(13)& 1.76(11)  & 5.1(4)   \\ \hline
18.3          & 0.32(5) & 1.26(11)& 1.46(23)  & 5.8(5)   \\ \hline
\end{tabular}}
\caption{
Effective screening masses from eq.\ \protect (\ref{scrmom}).
}
\label{tscreenmom}
\end{table}

We also like to mention our experience with a momentum space
analysis of the correlation functions, which has been popular
in numerical studies of the Higgs-Yukawa sector of the Standard Model.
This analysis is based on the Fourier transform of $G_{H,W}(d)$,
\be
G_{H,W}(p) = \sum_d e^{-ipd}\, G_{H,W}(d).
\label{FT}
\ee
Here $p$ is one of the lattice momenta, $ap=2\pi n/N$, $n=0,1,\cdots, N-1$.
Within statistical errors these are real functions of $p$.
We deduced effective screening masses $m_{\rm eff}$ from the expansion
\be
G(p)^{-1} = Z^{-1} [a^2 m_{\rm eff}^2 + p^2 + O(p^4)],
\label{scrmom}
\ee
by fitting $G^{-1}$ as a function of $\hat{p}^2\equiv 2-2\cos ap$ to a
straight line,
using two of the lower $p$ values ($n=ap/(2\pi/N) = 0,1$ and 1,2).
The resulting effective screening masses (cf.\ Table \ref{tscreenmom})
tend to be significantly larger than the true screening masses above,
especially for $W$ in the plasma phase, for which $G^{-1}(p)$ shows
strong curvature.
(Fitting more data points to a quadratic function had no significant effect
on the results.)
The effective masses are expected to be a good approximation to the
screening masses only in case of pole dominance of $G(p)$, which evidently
is not the case for $W$ in the plasma phase.

It is known that in the Higgs phase the simple fields (\ref{OW}, \ref{OH})
which we use
perform reasonably well in creating dominantly the $H$ and $W$
out of the vacuum, but in the plasma phase they easily
create also excited states with the same quantum numbers as $H$
and $W$ (for a recent study see \cite{Phiea96}). The excited state
contribution weakens of course pole dominance.
To do better one has to use more sophisticated
observables instead of the simple $H_{\vecx}$ and $W_{k\vecx}^{\al}$,
e.g.~the smeared fields used in \cite{Phiea96}.

%\section{Results for the plasmon correlators}
\section{Plasmon correlators in the Higgs phase}

\begin{figure}
\epsfxsize 150mm
%\centerline{\epsfbox{sm.all_hplas.eps}}
\centerline{\epsfbox{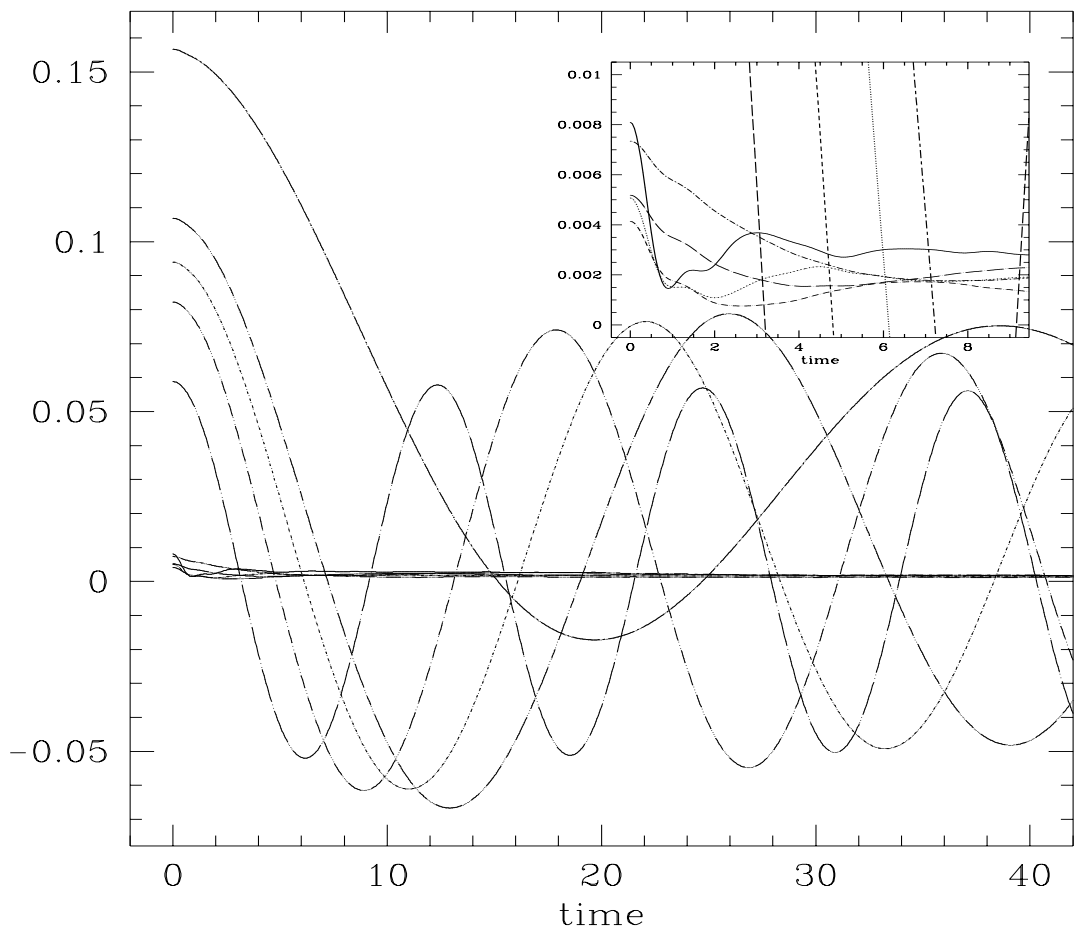}}
\caption{Scan of $C_H(t)$ from the $24^3$ data.
The first minima in $t > 5$ are in order of decreasing $\btb = 20$,
14, 13, 12.5, 12.
The lines in the middle of the $C_H$ plot are for $\btb < 12$,
shown enlarged in the insert.
}
\label{fscanh}
\end{figure}

\begin{figure}
\epsfxsize 150mm
\centerline{\epsfbox{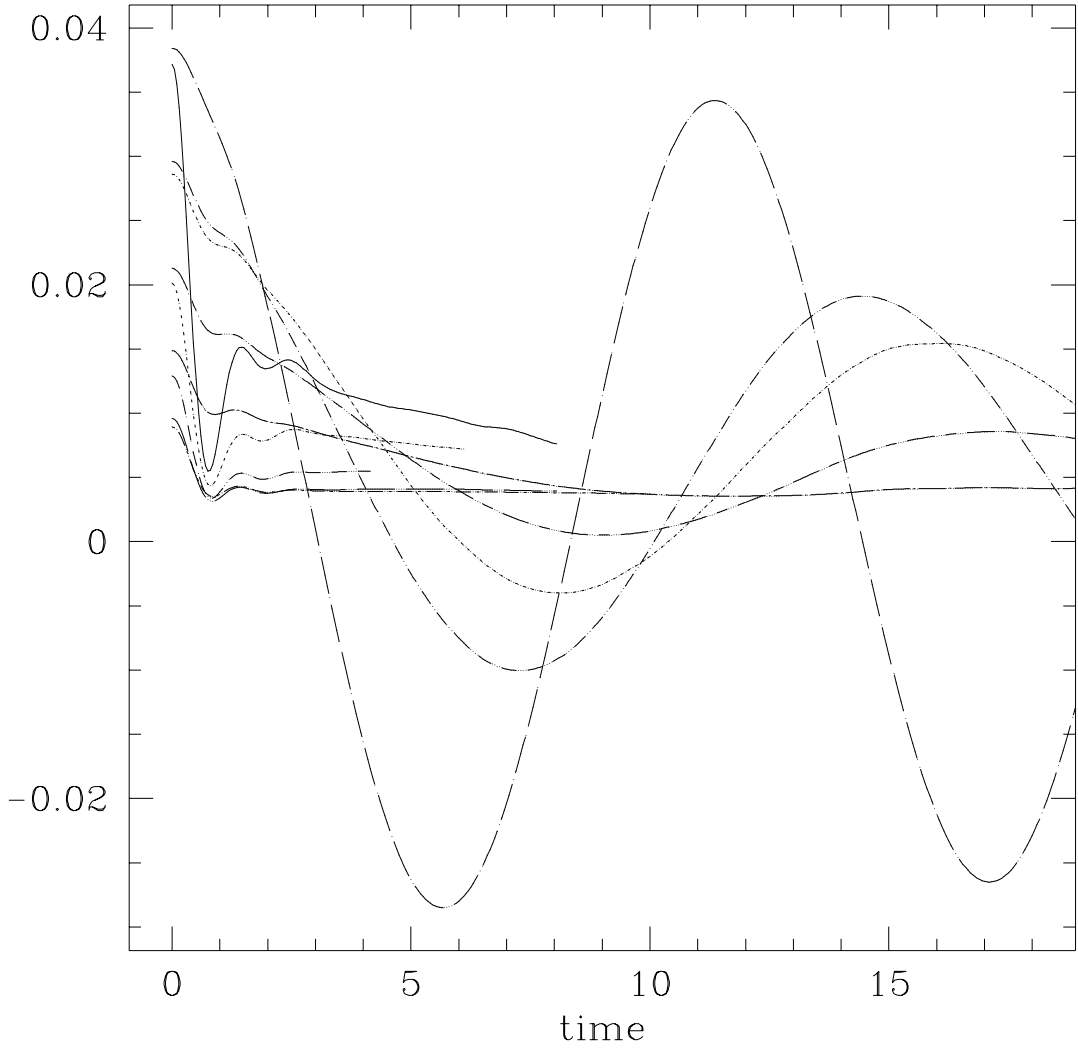}}
\caption{Scan of $C_W(t)$ from the $24^3$ data.
The first minima in $t > 5$ are in order of decreasing $\btb = 20$,
14, 13, 12.5, 12. The lines ending before $t=10$ are for $\btb < 12$;
bottom to top ($t = 4$): $\btb=11$ (touching the minimum of $\btb=12$),
10 (nearly coinciding with 11), 8, 6 and 4.
}
\label{fscanw}
\end{figure}

The results in the previous section for the screening properties of
$H$ and $W$ give us a feeling how well we are doing in a familiar
situation, in the theoretical framework of dimensional
reduction. We now turn to the time dependent correlation functions,
for which we are on less established ground.
{}Figs.\ \ref{fscanh} and \ref{fscanw} show an initial scan
of the autocorrelation
functions $C_H(t)$ and $C_W(t)$ for various $\btb$ at fixed
$\vb^2 = \vb^2_c(\btb=12)$ on the $24^3$ lattice.
The horizontal time scale is in lattice units
(i.e.\  (number of time integration steps) $\times$ (step size)
as it appears in the computer code). The time in lattice units can be
converted to physical units by multiplication with e.g.\ the
temperature in lattice units.
We see the period of the oscillations increasing as we approach
the transition at $\btb = 12$ from above
(from the Higgs side)\footnote{The
curves at $\btb=12$ belong clearly to the Higgs phase sequence,
which suggests that $\btb_c$ is actually slightly smaller that 12.}.

In the $H$ case (Fig.\ \ref{fscanh}) there is a drastic change of
behavior below $\btb=12$ (the plasma side), where the signal is
much smaller.  As the insert shows the data here is very noisy,
although we think the very short time ($t<2$) data is still
relevant, because short times allow for more microcanonical
averaging. Notice the oscillations in this very short time region.

For $W$ (Fig.\ \ref{fscanw}) the transition appears more gradual.  As
$\btb$ is lowered down from 20, small oscillations in the small time
region ($t<2$) appear already in the Higgs phase.
When $\btb$ is
decreased further below 12 the initial oscillations increase somewhat
and so does the slope afterwards, which is nearly zero for $\btb=11$, 10, 8.
In this case we have stopped the
drawing of the
curves at early times because the
statistical noise would be overwhelming.

To extract a plasmon frequency and damping rate we fit the data in the
Higgs phase to the simple asymptotic form
\be
R e^{-\gm t} \cos (\om t + \al) + C.
\label{cform}
\ee
For a free scalar field we would have $R=T/m^2$, $\gm=C=0$. With interactions
the form (\ref{cform}) is expected to be a good approximation in
case of small damping \cite{Boyaea95}.
It turns out that the phase $\al$ is not really needed and
may be set equal to zero.
In the plasma phase the signal of our $24^3$ data
is too noisy for a quantitative analysis.
For the Higgs phase the $24^3$ data led to the plasmon parameters shown in
Tables \ref{tplasfreq} and \ref{tdamping}.
The errors are based on jackknifing with respect to the initial
conditions. The damping rate $\gm$ is sensitive to the beginning of
the fitting range, the fits above started roughly from the third  maximum
of $C_H(t)$
(the first maximum is at $t=0$).
\begin{table}[h]
\centerline{
\begin{tabular}{||c|c|c|c|c|c|c||} \hline
\multicolumn{1}{||c|}{$\bar{\beta}$} &
\multicolumn{1}{c|}{$N$} &
\multicolumn{1}{c|}{$T/T_c$} &
\multicolumn{1}{c|}{$a\om_{H}$} &
\multicolumn{1}{c|}{$a\om_{W}$} &
\multicolumn{1}{c|}{$\om_{H}/g^2 T$} &
\multicolumn{1}{c||}{$\om_{W}/g^2 T$}
\\ \hline
20       &24 &0.41 & 0.508(2)& 0.551(2)& 2.54(1) & 2.76(1) \\ \hline
14       &24 &0.74 & 0.350(4)& 0.430(6)& 1.23(2) & 1.51(2) \\ \hline
21.7     &32 &0.78 & 0.210(2)& 0.296(2)& 1.14(1) & 1.61(1) \\ \hline
13       &24 & 0.85& 0.28(2) & 0.38(2) & 0.91(7) & 1.24(7) \\ \hline
$13'$    &20 &0.85 & 0.284(5)& 0.382(5)& 0.92(2) & 1.24(2) \\ \hline
12.5     &24 & 0.92& 0.24(1) & 0.35(2) &  0.75(3)& 1.09(6) \\ \hline
12       &24 & 1   & 0.16(1) & ---     & 0.48(3) & --- \\ \hline
$11'$    &20 & 1.23& 0.33(3) & ---     & 0.91(8) & --- \\ \hline
$11^\ast$&20 & 1.52& 0.44(3) & ---     & 1.21(8) & --- \\ \hline
18.3     & 32& 1.52& 0.23(2) & ---     & 1.05(9) & --- \\ \hline
% & & & & & & & & & \\ \hline
\end{tabular}}
\caption{Results for the plasmon frequencies.}
\label{tplasfreq}
\end{table}
\begin{table}[h]
\centerline{
\begin{tabular}{||c|c|c|c|c|c|c||} \hline
\multicolumn{1}{||c|}{$\bar{\beta}$} &
\multicolumn{1}{c|}{$N$} &
\multicolumn{1}{c|}{$T/T_c$} &
\multicolumn{1}{c|}{$a\gm_{H}$} &
\multicolumn{1}{c|}{$a\gm_{W}$} &
\multicolumn{1}{c|}{$\gm_{H}/g^2 T$} &
\multicolumn{1}{c||}{$\gm_{W}/g^2 T$}
\\ \hline
20         & 24 & 0.41 & 0.0012(19)& 0.008(4)  & 0.006(10) & 0.04(2) \\ \hline
14         & 24 & 0.74 & 0.0045(31)& 0.03(1)   & 0.016(11) & 0.105(35) \\ \hline
21.7       & 32 & 0.78 & 0.0052(8) & 0.041(8)  & 0.028(4)  & 0.22(4) \\ \hline
13         & 24 & 0.85 & 0.013(29) & 0.04(5)   & 0.042(94) & 0.13(16) \\ \hline
$13'$      & 20 & 0.85 & 0.011(5)  & 0.08(2)   & 0.036(16) & 0.26(7) \\ \hline
12.5       & 24 & 0.92 & 0.012(8)  & 0.09(9)   & 0.038(25) & 0.28(28) \\ \hline
12         & 24 & 1    & 0.03(2)   & --        & 0.09(6)   & --- \\ \hline
$11'$      & 20 & 1.23 & 0.079(61) & 0.0116(17)& 0.22(17)  & 0.032(5) \\ \hline
$11^\ast$  & 20 & 1.52 & 0.053(55) & 0.0111(6) & 0.15(15)  & 0.031(2) \\ \hline
18.3       & 32 & 1.52 & 0.036(40) & 0.0062(8) & 0.16(18)  & 0.028(4) \\ \hline
% & & & & & & & & & \\ \hline
\end{tabular}}
\caption{Results for the damping rates.}
\label{tdamping}
\end{table}

\begin{figure}
\epsfxsize 150mm
\centerline{\epsfbox{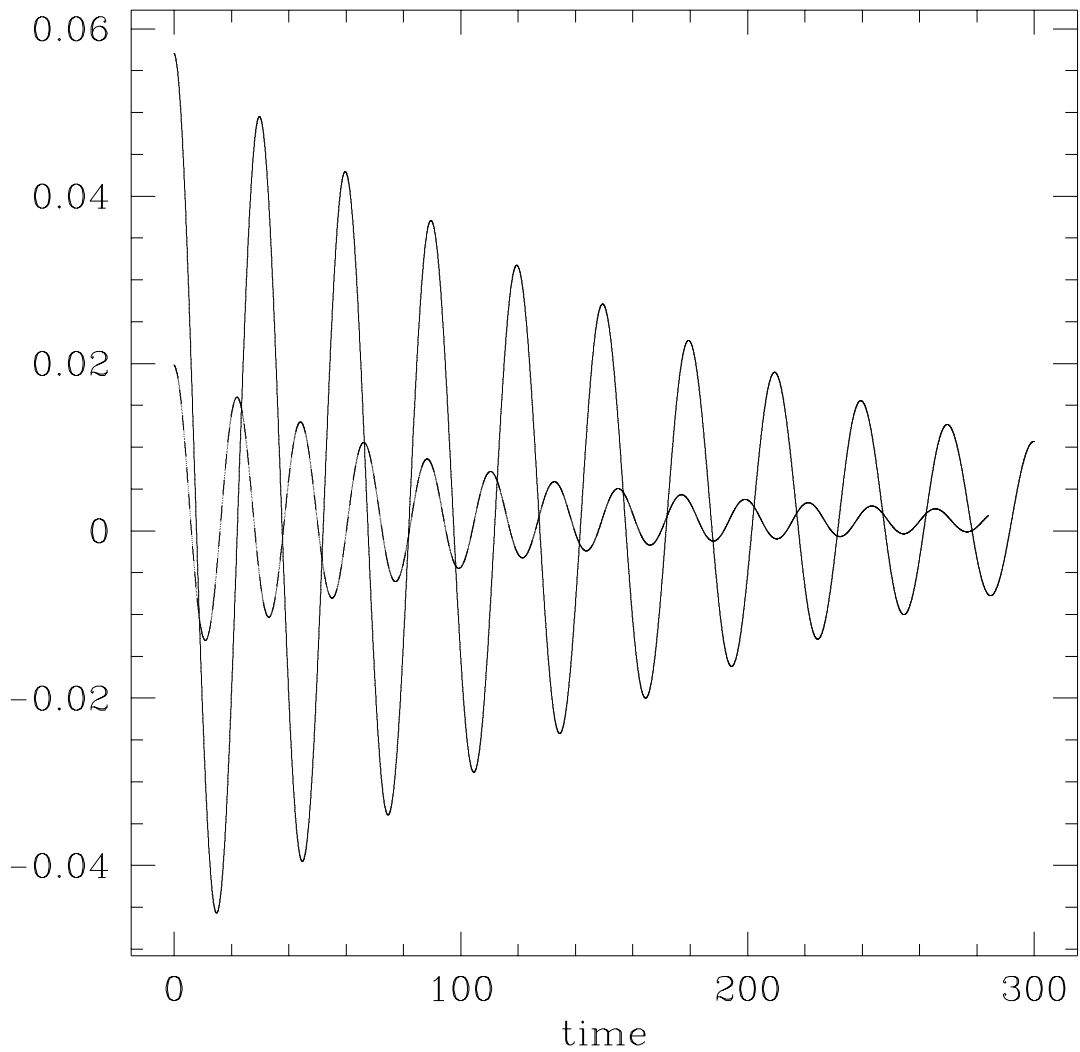}}
\caption{The autocorrelator $C_H(t)$ in the Higgs phase for
$\btb = 13'$ (small amplitude) and 21.7 (large amplitude).}
\label{fchh}
\end{figure}

\begin{figure}
\epsfxsize 150mm
\centerline{\epsfbox{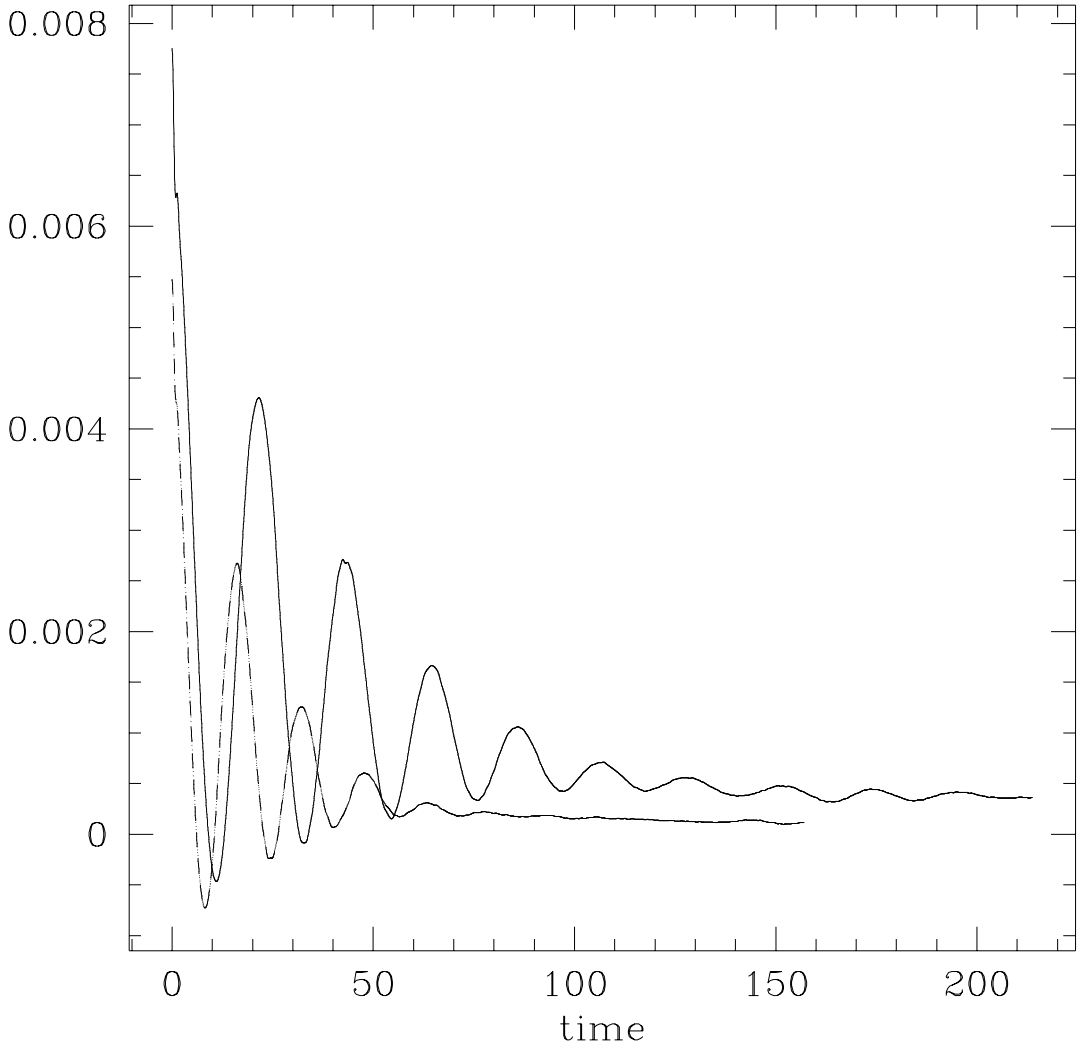}}
\caption{The autocorrelator $C_W(t)$ in the Higgs phase for
$\btb = 13'$ (lower tail) and 21.7 (upper tail). }
\label{fcwh}
\end{figure}

We increased the numerical effort on the plasma phase
parameters values $\btb=11'$, $11^*$ and 18.3
and did similar runs for $13'$ and 21.7.
{}Fig.~\ref{fchh} shows the autocorrelation function $C_H(t)$ in the Higgs
phase for $\btb=13'$ and 21.7.
A fit to (\ref{cform}) starting roughly at the third maximum
led to the plasmon parameters shown in Tables
\ref{tplasfreq} and \ref{tdamping}.
Figure \ref{fcwh} shows the $W$-autocorrelation functions in the Higgs
phase. It is clear that the $W$ damping rate is considerably
larger than for $H$ and to avoid running into noise
we had to lower the starting time of the fit to (\ref{cform})
down to roughly the second maximum.
The constant $C$ of the fits is non-zero.
However, we have seen the autocorrelators to tend to zero at very large times.
The results in physical units are shown in Figs.\ \ref{fplas} and \ref{fdamp}.
{}From the smooth adjustment of the $\btb = 21.7$ ($32^3$) data with
the $\btb=13'$ and 13 ($20^3$ and $24^3$) data it follows
that the plasmon frequencies and damping rates are nearly lattice
spacing independent.

However, 
the $32^3$ data point in Fig.\ \ref{fplas}
for the $W$ plasmon frequency 
at $T/T_c=0.78$ does look somewhat high
compared to the trend of the $24^3$ data. 
Interpolating the $24^3$ data linearly between $T/T_c=0.74$ 
and 0.85 gives us a hypothetical data point $\om_W/g^2 T = 1.41(3)$ at 
$T/T_c = 0.78$, $\btb = 13.6$.
The $32^3$ 
%($\btb = 21.7$) 
data point at $T/T_c = 0.78$, 1.61(1), is a factor 1.14 higher. 
This is much less than the 
corresponding ratio of lattice spacings, 21.7/13.6 = 1.60, and in
the previous version of this paper we ignored this 14\% discrepancy 
as due to a possible systematic error in our data analysis. Now we
shall take it seriously and try to interpret it as an expression of 
lattice spacing dependence. 
As a rough model, let us assume that $\om_H^2 = 2\lm v^2$,
$\om_W^2 = g^2 v^2/4 + \tilde\om^2$, where $v^2$ is 
%a renormalized expectation value of the Higgs field, 
simply defined by
the equation $v^2 = \om_H^2/2\lm$, and $\tilde\om^2$ is an additional
contribution to the $W$-plasmon frequency.
Since the Higgs parameters are tuned according to dimensional
reduction we assume that $\om_H^2$ is independent of $a$.
This is consistent with the trend of the data in Fig.\ \ref{fplas}.
With $\lm/g^2 = 1/8$ we get $\om_W^2 = \om_H^2 + \tilde\om^2$.
Hard thermal loop arguments \cite{Ar97} lead to the conclusion
that there is a divergent component in $\om_W^2$,
given by 
$0.08606 C_A g^2 T/a$, with $C_A = 2$ for SU(2). 
We have checked in a one loop calculation perturbative calculation
on the lattice that a scalar gives the same with $C_A \to 1/2$.
Let us now assume $\tilde\om^2$ to be this divergent contribution,
and write $\tilde\om^2 = \dl g^2 T/a$. Then $\dl = (2+1/2)\, 0.08606 = 0.2151$.
Denoting the interpolated $\btb = 13.6$
data by the subscript 1 and the $\btb=21.7$ 
data without a subscript,
we have with  $\om_H/g^2 T = 1.14$, $a\om_H = 0.210$ 
(cf Table \ref{tplasfreq})
and  $a_1\om_H = a\om_H*(a_1/a) = 0.210*(21.7/13.6)$,
\be
\frac{\om_W}{\om_{1W}} = 
\sqrt{\frac{\om_H^2 + \tilde\om^2}{\om_H^2 + \tilde\om_1^2}}
=
\sqrt{\frac{1+\dl (g^2 T/\om_H)/(a\om_H)}{1+\dl (g^2 T/\om_H)/(a_1\om_H)}}
= \sqrt{\frac{1 + 4.18\, \dl}{1+2.62\, \dl}} = 1.10,
\ee
for $\dl = 0.2151$. 
This is somewhat smaller than the `measured' ratio 
$\om_W/\om_{1W} = 1.61/1.41 = 1.14$.
Using this as input and solving for $\dl$ would give $\dl = 0.38$.
Notice that the $\dl$-terms in the last square root above are
not small compared to one.
In this interpretation all the difference 
between $\om_H$ and $\om_W$ is ascribed to the $\dl$-contribution,
which is probably not the whole story in a more detailed
perturbative calculation, but which works reasonably well. 
For example, $\om_W/\om_H - 1 = 0.41$
for the 21.7 data, while $\sqrt{1+4.18\, \dl} - 1 = 0.38$ for $\dl = 0.2151$.
Furthermore, for $\btb = 12.5$, 13, 14, 20 the predicition
for $\om_W/\om_H -1$ is 0.48, 0.36, 0.22, 0.080, respectively,
to be compared with the data 0.46, 0.36, 0.23, 0.085. 
The above interpretation suggests 
that the approximate lattice spacing independence
of our data for the plasmon frequencies
is fortuitous and that it may well turn into a 
clear $1/\sqrt{a}$ behavior at much smaller lattice spacings.
 
\section{Autocorrelators in the plasma phase}

\begin{figure}
\epsfxsize 150mm
%\centerline{\epsfbox{sm.h_symm.eps}}
\centerline{\epsfxsize=75mm\epsfbox{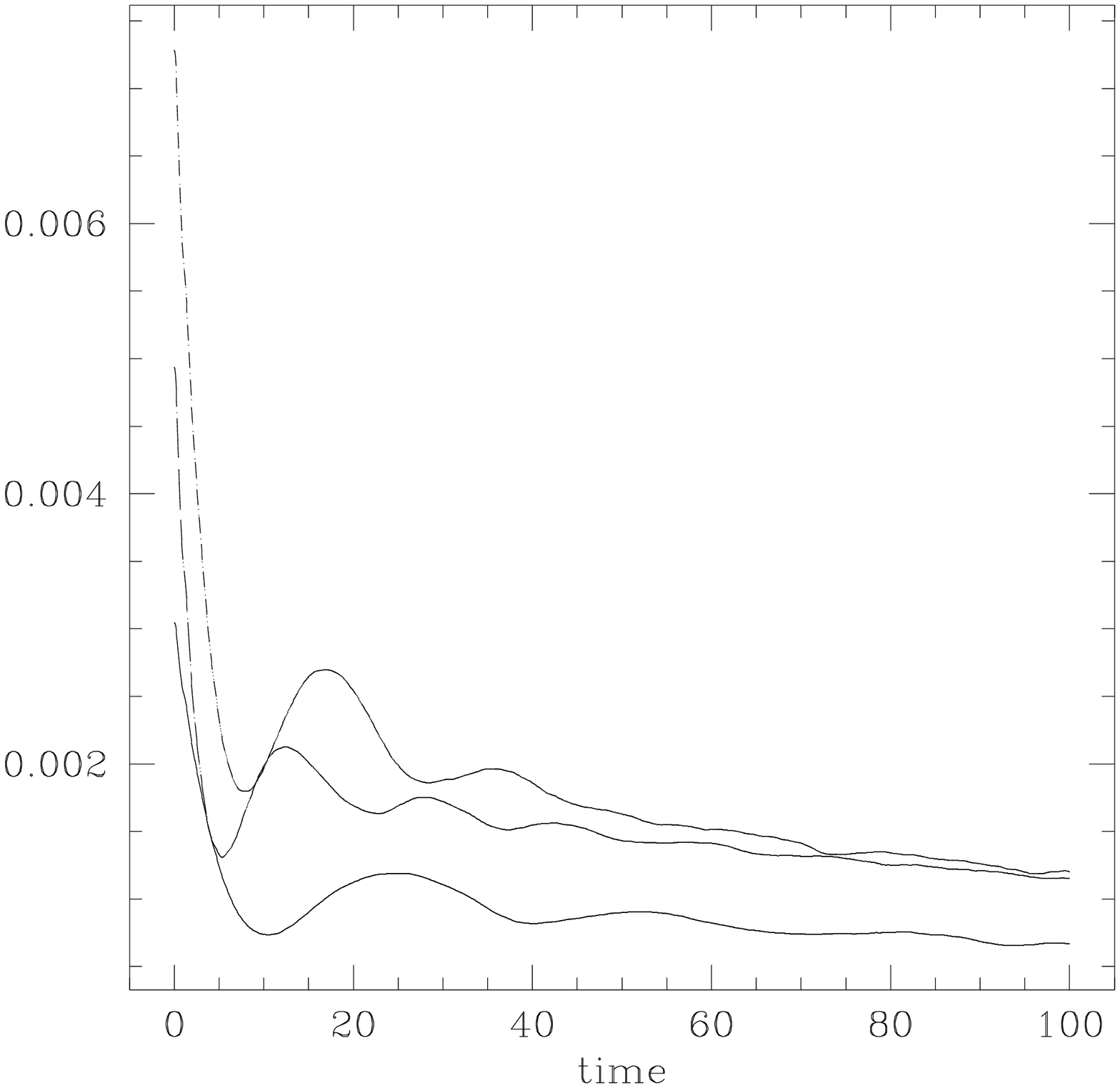}
            \epsfxsize=75mm\epsfbox{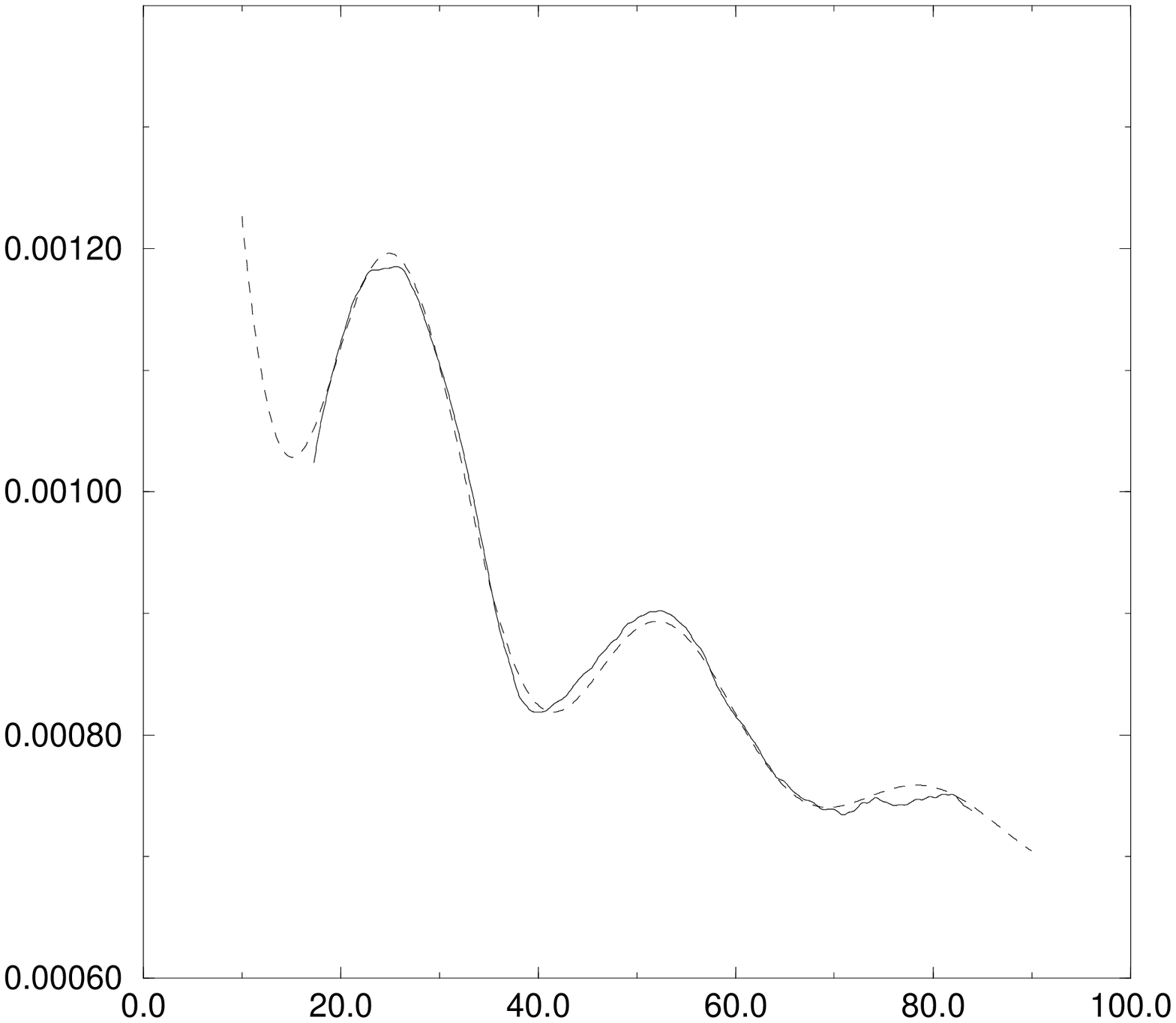}
}
\caption{Left: $C_H(t)$ in the plasma phase;
$\btb = 18.3$, $11^\ast$ and $11'$ in ascending order.
Right: fit (dashed curve) to the data at $\btb=18.3$;
the fitting region corresponds to the data shown.
}
\label{fchc}
\end{figure}

\begin{figure}
\epsfxsize 150mm
\centerline{\epsfxsize=75mm\epsfbox{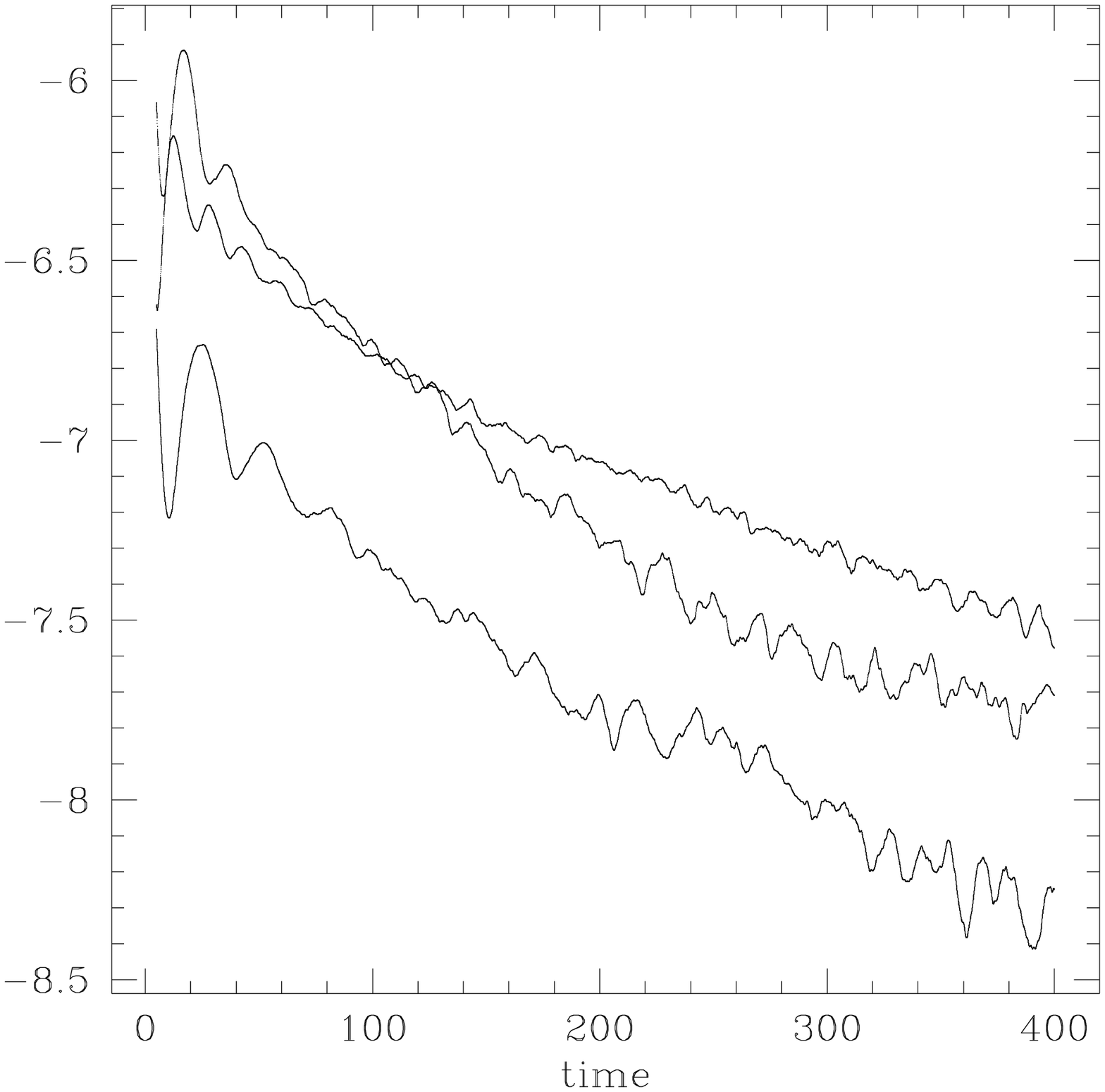}
            \epsfxsize=75mm\epsfbox{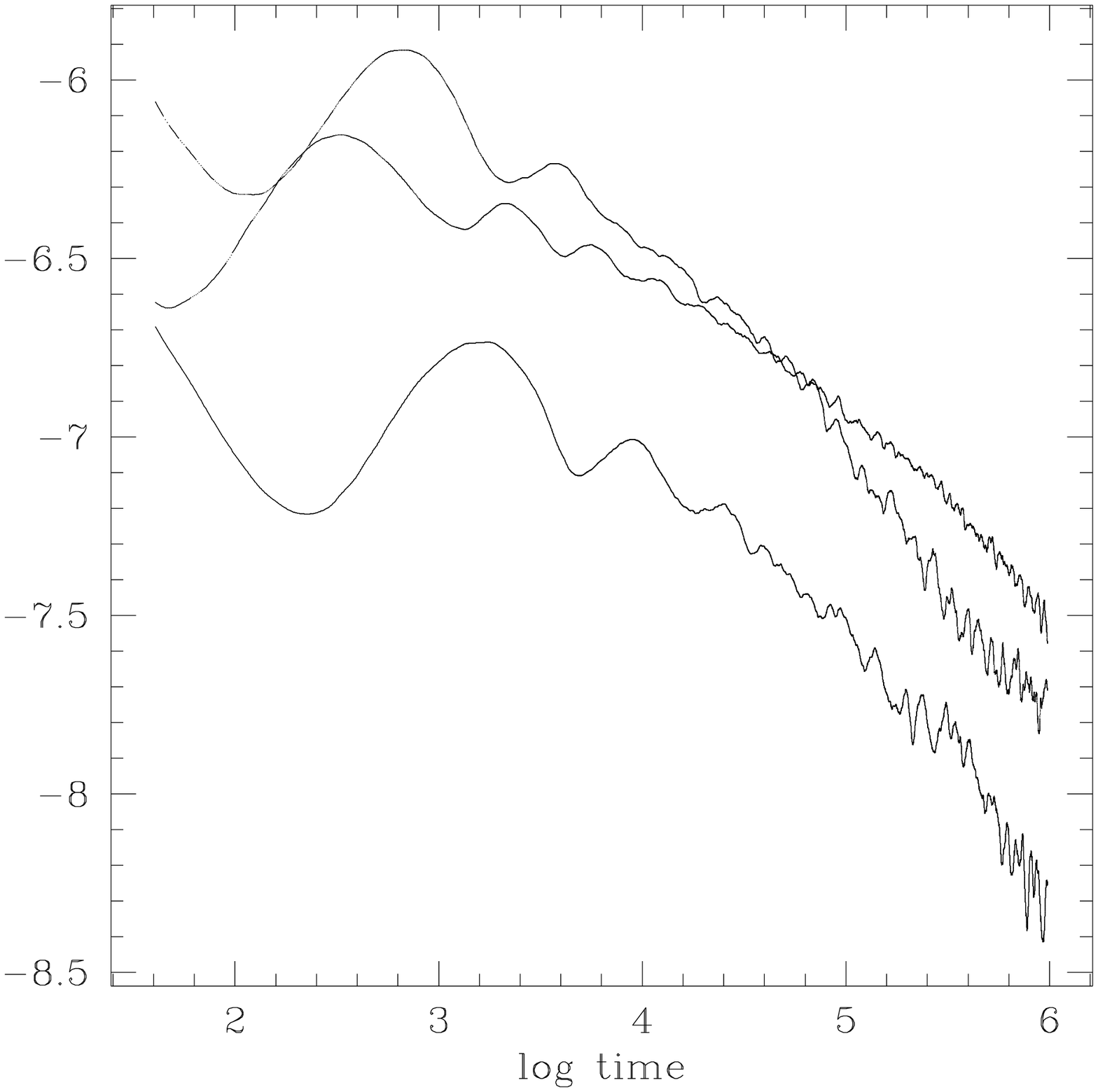}}
\caption{$\ln C_H(t)$ versus $t$ (left) and $\ln C_H(t)$ versus $\ln t$
(right) in the plasma phase.
}
\label{fchcll}
\end{figure}

In the plasma phase the behavior is qualitatively different.
{}Fig.~\ref{fchc} shows that $C_H(t)$ still has roughly the form
(\ref{cform}), but with an average $C$ that gets significantly smaller
in a period of oscillation. A reasonable fit to the data could be
obtained by letting $C \to C t^{-\dl}$ in a region starting a little
before the second maximum to times where the oscillations
appear to have died out, judging by eye.
An example of such a fit is shown in figure \ref{fchc}.
The large statistical errors for $C_H$ in the Higgs phase
show up in the lack of smoothness in the data and the resulting
jackknife error on $\gm_H$ is very large.
The results are in Tables \ref{tplasfreq}, \ref{tdamping} and
Figs.\ \ref{fplas} and \ref{fdamp}.
The fitted power $\dl$ is about
0.2 -- 0.4. For larger times ($t > 100$ for $11^*$) the power $\dl$
appears to increase, as shown in the double logarithmic plot Fig.\
\ref{fchcll}, which also shows that there is no good evidence for
exponential behavior.  Asymptotically one would expect the $t^{-3/2}$
behavior familiar from zero temperature \cite{Boyaea95}.
 
\begin{figure}
\epsfxsize 100mm
\centerline{\epsfbox{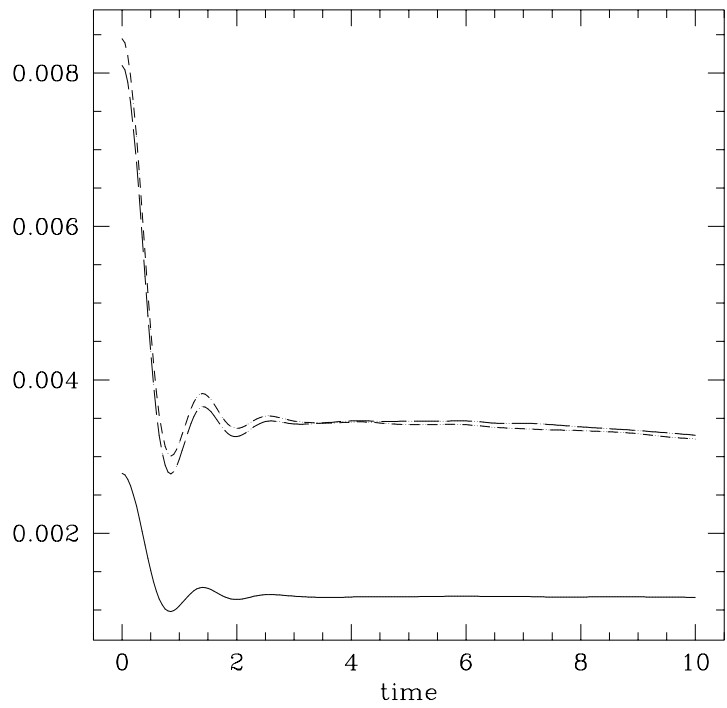}}
\caption{Short time behavior of $C_W(t)$ in the plasma phase.
Top to bottom ($t\approx 1$):
$\btb = 11'$, $11^\ast$ and 18.3. }
\label{fcwc1}
\end{figure}

The $W$ autocorrelator behaves very
differently. Fig.~\ref{fcwc1} shows what looks like $C_H(t)$ in
Fig.~\ref{fchc}, except that the time scale has shrunk by an order of
magnitude. The periods of the oscillations are of order 1, which
suggests $\om$'s of order $2\pi$ and also the $\gm$'s appear to be
very large in lattice units.  There is clearly no scaling in this time
interval since the period of the oscillation hardly changes as $\btb$
increases. We recall that the oscillations start already in the Higgs
phase quite far from the phase transition, as mentioned above after
introducing Fig.\ \ref{fscanw}.  They are not analogous to those of
$C_H(t)$ in Fig.~\ref{fchc}, but of $C_H(t)$ in the insert in Fig.\
\ref{fscanh}.  The small time oscillations correspond to energies of
the order of the cutoff and must be considered as 
dominated by lattice artefacts.

\begin{figure}
\epsfxsize 150mm
\centerline{\epsfbox{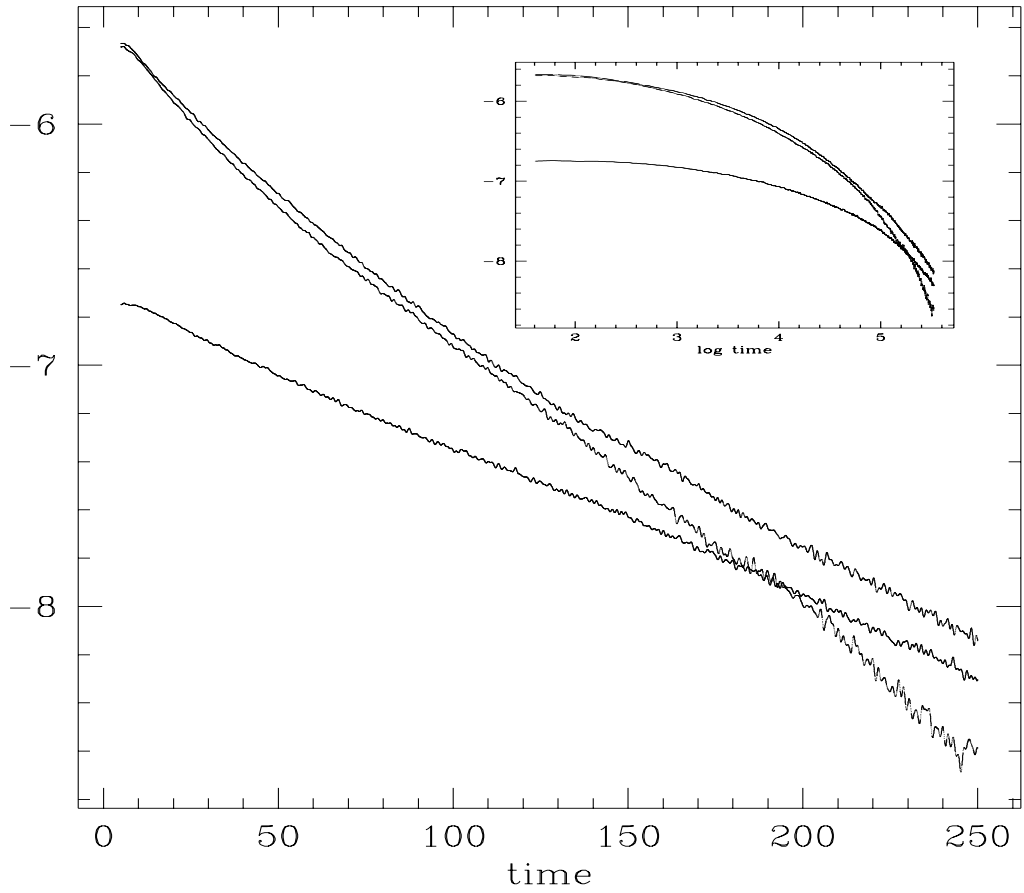}}
\caption{Plot of $\ln C_W(t)$ in the plasma phase for larger times;
$\btb = 11'$ (top), $11^\ast$ (middle) and 18.3 (bottom). The insert
shows the same time interval on a double logarithmic plot.}
\label{fcwc2}
\end{figure}

Fig.~\ref{fcwc2} shows the larger time region on a logarithmic
scale. We see good evidence for exponential behavior.  A power
behavior appears to be excluded in the time range shown (cf.\
insert). The drop of $C_W(t)$ can be fitted to an exponential with a
damping rate in physical units that is lattice spacing independent,
within errors, as shown in Table \ref{tdamping} and Fig.\
\ref{tdamping}.  On the other hand, the slopes of $\btb=11'$ and
$11^{\ast}$ are equal within errors which suggests that the $W$
dynamics is practically independent of the scalar field dynamics in
the plasma phase.  A similar effect has been noticed for the
screening lengths, see e.g.\ \cite{Phiea96}.  There appear to be no
oscillations in the exponential regime.  At very large times we found
irregular oscillations with a period of order 600 (18.3 data),
indicating a frequency of order 0.01, but the statistics is
insufficient to be able to say anything more definite.  

Ignoring this hint of oscillations at large times as being noise,
it is interesting to explore an interpretation in terms of overdamping
such as discussed in \cite{ArSoYa97,Ar97}. Suppose the $W$ mode 
satisfies an equation of the form 
$(\partial_t^2 + \Gm \partial_t + M^2)C_W(t) = 0$ for $t > 0$,
with some basic frequency $M$ and damping rate $\Gm$.
Substituting $\exp(-i\om t)$ gives the equation
$-\om^2 -i\Gm \om + M^2 = 0$. For 
small $\Gm^2\ll M^2$ this leads to damped oscillations, 
but for large $\Gm^2\gg M^2$ the system is overdamped, 
in the sense that the solutions $\om$ are purely imaginary and no oscillations
occur: $\om_1 \approx -i\Gm$, $\om_2 \approx -iM^2 /\Gm$. 
Identifying the smaller $|\om_2|$ with our measured damping rate $\gm_W$,
we can try to interpret $C_W(t)$ for $\btb = 18.3$ with $a\Gm =  O(0.5)$ 
(qualitatively representing the strong damping in Fig.\ \ref{fcwc1},
while ignoring the oscillations)
and $aM = a\sqrt{\gm_W \Gm} = \sqrt{0.0062\, a\Gm} = O(0.06)$. 
However, since the above equation for $C_W(t)$
is supposed to hold under the condition
$|\om|^2 \ll M^2$, it seems best to discard $\om_1$ altogether.
It is interesting that $\gm_W$ is inversely proportional to $\Gm$.

\section{Discussion}
\begin{figure}
\epsfxsize 150mm
\centerline{\epsfbox{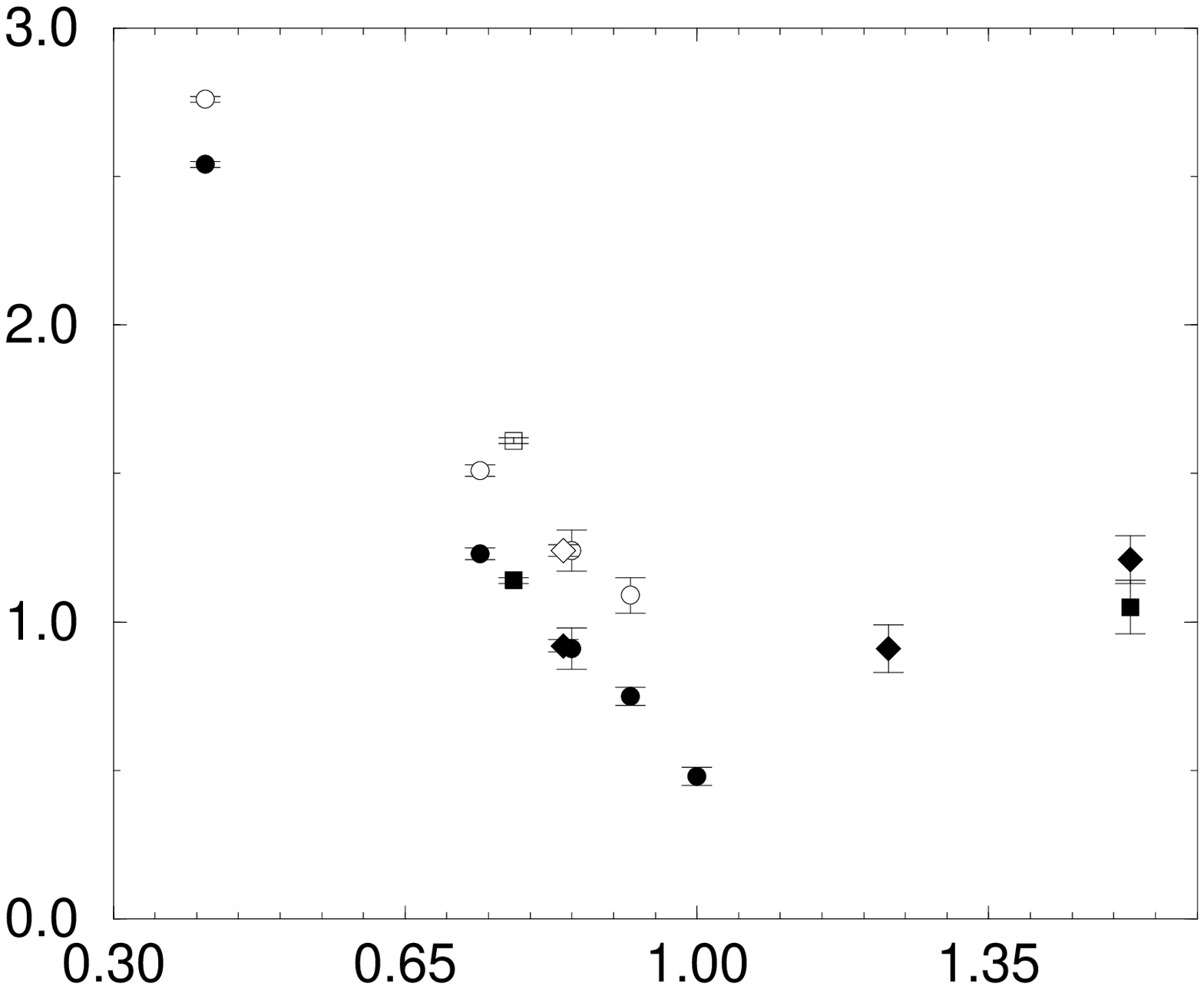}}
\caption{Plasmon frequencies $\om/g^2 T$
versus $T/T_c$. Solid symbols correspond to $H$, open symbols to $W$;
circle: $24^3$ data; diamond: $20^3$ data; square: $32^3$ data.
}
\label{fplas}
\end{figure}

\begin{figure}
\epsfxsize 150mm
\centerline{\epsfbox{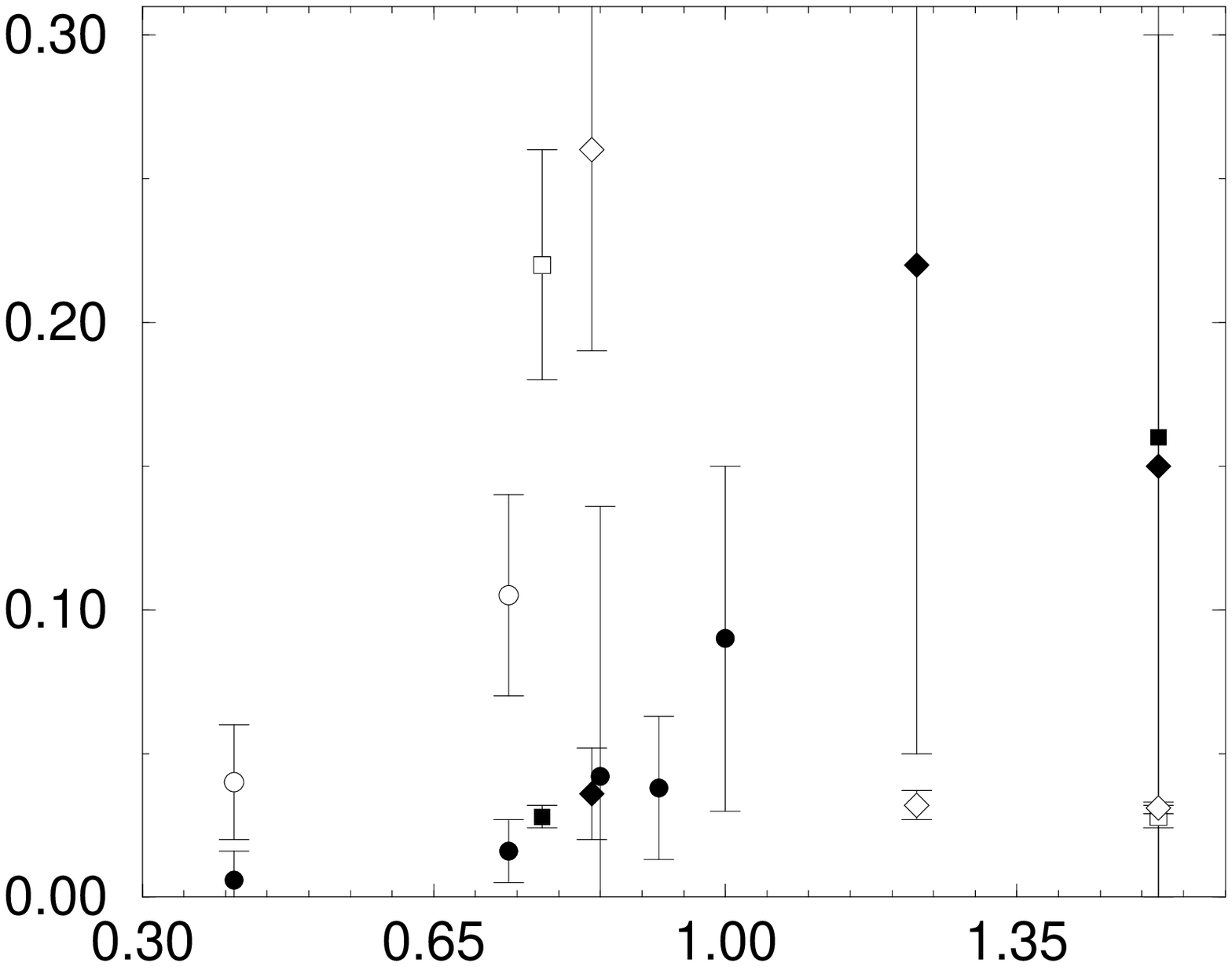}}
\caption{Damping rates $\gm/g^2 T$
versus $T/T_c$. Solid symbols correspond to $H$, open symbols to $W$;
circle: $24^3$ data; diamond: $20^3$ data; square: $32^3$ data.
}
\label{fdamp}
\end{figure}

We have learned that numerical computations of real time correlation
functions in hot SU(2)-Higgs theory are feasible in the classical
approximation. In this exploratory study we have not indicated
statistical errors in our plots of the correlation functions.  In
most cases these are quite reasonable, but in some they would swamp
the data lines, e.g.\ for $H$ in the plasma phase.  However, we
believe the jackknife errors given in Tables \ref{tplasfreq},
\ref{tdamping} and Figs.\ \ref{fplas}, \ref{fdamp} are mostly
realistic and they can be improved with a reasonable increase of
computational effort. The errors on $\gm_H$
appear to be overestimated, but this can only be clarified by
increasing statistics.
Systematic errors are harder to assess.  It is one thing to compute an
autocorrelation function, and quite another to extract plasmon
parameters from these when the damping rate is relatively large. But
then the very relevance of such concepts is questionable in case of
strong damping.

The results for the $H,W$ plasmon frequencies and damping rates 
appear to be roughly independent of the lattice spacing $a$. 
We found, however, that a small effect in the 
$W$ frequency at $T/T_c = 0.78$ could be interpreted as 
compatible with the expected $1/\sqrt{a}$ divergence \cite{Ar97}.
The approximate $a$-independence of the $H$ frequency may due
to the fact that Higgs screening mass is already tuned by the
counterterms based on dimensional reduction (for the $W$
there is no room for such tuning within the classical
form of the hamiltonian).

As argued in the Introduction, comparison with analytical results
in the quantum theory is to be done in the Higgs phase.
The numerical damping rates in the Higgs phase compare reasonably well:
the $\btb=21.7$ ($T/T_c = 0.78$) data
give
$\gm_W/g^2 T = 0.22(4)$ and $\gm_H/g^2 T = 0.028(4)$, to be compared with
$0.176$ and $\approx 0.02$, respectively \cite{BraPi90,Birea96}.
Of course, these values will depend
on the distance from the transition at $T_c$
(cf.\ Fig.\ \ref{fplas}).
The approximate lattice spacing independence of the damping rates may 
express an insensitivity to the magnitude of the plasmon frequency. 
This is the case in the quantum theory where the plasmon frequency
is of order $gT$ and the damping rate of order $g^2 T$, which is
the primary scale of the classical theory. We note also that the
perturbative values for the quantum and classical plasmon frequencies 
for the gauge field are in fact
not very different in our simulation: for SU(2) gauge theory we have   
$a\om_W^{\rm qu}= 2\sqrt{2}/\btb$ and 
$a\om_W^{\rm cl} = 0.8297/\sqrt{\btb}$ \cite{Ar97}, which are equal
for $\btb =11.6$.

The damping rate associated with $H$ increases substantially when the 
temperature is raised above $T_c$, while the rate associated with
$W$ drops dramatically. Both damping rates
appear to be roughly temperature independent in the plasma phase,
although the errors on $\gm_H$ are much too large for a definite conclusion.
The data for the $W$ correlation function in the plasma phase
suggest overdamping, with a rather small damping rate. For a proper
interpretation we may have to do better than just a pole approximation
to the correlation function.
It is interesting that a similar phenomenon was found 
recently in a simulation of correlators of SU(2) gauge fields, 
transformed to Coulomb gauge \cite{AmKra97}.
We stress again that in the plasma phase the `plasmon' frequencies
and damping rates of the composite $H$ and $W$ fields
should not be identified with those of the elementary Higgs
and gauge fields.
\\[2mm]

\noindent{\bf Acknowledgement} \\
The numerical simulations made use of
the code \cite{Kra95} kindly provided to us by Alex Krasnitz.
We thank Gert Aarts, Peter Arnold and Alex Krasnitz for useful remarks.
This work is supported by
%the Stichting voor Fundamenteel Onderzoek der Materie (FOM)
FOM and
%the Stichting Nationale Computerfaciliteiten (NCF),
NCF, with financial support from
%the Nederlandse Organisatie voor Wetenschappelijk Onderzoek (NWO).
NWO.

%\input{plasbib}

%\input{figscreen}
%\begin{figure}[p]

%\input{figscanh}
%\begin{figure}[p]
%\input{figscanw}
%\begin{figure}[p]

%\input{figchh}
%\begin{figure}[h]
%\begin{figure}[p]
%\input{figcwh}
%\begin{figure}[h]
%\begin{figure}[p]

%\input{figfreqdamp}
%\begin{figure}[p]
%
%\begin{figure}[p]

%
%\input{figchcll}
%\begin{figure}[h]
%\begin{figure}[p]

%\input{figcwc1}
%\begin{figure}[h]
%\begin{figure}[p]
%
%\input{figcwc2}
%\begin{figure}[h]
%\begin{figure}[p]

\end{document}